\documentclass[twocolumn,showpacs,%
amsmath,amssymb]{revtex4}

\usepackage{subfigure}
\usepackage{amsmath,amssymb}
\usepackage{graphicx}
\usepackage{rotating}
\usepackage{bm}
\usepackage{color}

\definecolor{blue}{rgb}{0,0,1}

\definecolor{green}{rgb}{0,1,0}

\definecolor{red}{rgb}{1,0,0}

\begin{document}

\author{Holger Gies}
\author{Klaus Klingm\"uller}
\affiliation{Institut f\"ur Theoretische Physik, Philosophenweg 16, %
   69120 Heidelberg, Germany} 

\title{Casimir effect for curved geometries: PFA validity limits}

\begin{abstract}
  
  We compute Casimir interaction energies for the sphere-plate and
  cylinder-plate configuration induced by scalar-field fluctuations
  with Dirichlet boundary conditions. Based on a high-precision
  calculation using worldline numerics, we quantitatively determine
  the validity bounds of the proximity force approximation (PFA) on
  which the comparison between all corresponding experiments and
  theory are based. We observe the quantitative failure of the PFA on
  the 1\% level for a curvature parameter $a/R> 0.00755$. Even
  qualitatively, the PFA fails to predict reliably the correct sign of
  genuine Casimir curvature effects. We conclude that data analysis of
  future experiments aiming at a precision of 0.1\% must no longer be
  based on the PFA.

\end{abstract}


\maketitle


Measurements of the Casimir force \cite{Casimir:dh} have reached a
precision level of 1\%
\cite{Lamoreaux:1996wh,Mohideen:1998iz,Roy:1999dx,Ederth:2000,%
Chan:2001,Bressi:2002fr,Chen:2002}.  Further improvements are
currently aimed at with intense efforts, owing to the increasing
relevance of these quantum forces for nano- and micro-scale mechanical
systems; also, Casimir precision measurements play a major role in the
search for new sub-millimeter forces, resulting in important
constraints for new physics \cite{Bordag:1998nv,Long:1998dk,%
Mostepanenko:2001fx,Milton:2001np,Decca:2003td}.

On this level of precision, corrections owing to material properties,
thermal fluctuations and geometry dependencies have to be accounted
for
\cite{Klimtchiskaya:1999,Lambrecht:1999vd,Bezerra:2000,Bordag:2001qi}.
In order to reduce material corrections such as surface roughness and
finite conductivity which are difficult to control with high
precision, force measurements at larger surface separations up to the
micron range are intended. Though this implies stronger geometry
dependence, this latter effect is, in principle, under clean
theoretical control, since it follows directly from
quantum field theory \cite{Graham:2002xq}.

Straightforward computations of geometry dependencies are conceptually
complicated, since the relevant information is subtly encoded in the
fluctuation spectrum.  Analytic solutions can usually be found only
for highly symmetric geometries. This problem is particularly
prominent, since current and future precision measurements
predominantly rely on configurations involving curved surfaces, such
as a sphere above a plate. As a general recipe, the
proximity force approximation (PFA) \cite{pft1} has been the standard
tool for estimating curvature effects for non-planar geometries in all
experiments so far. The fact that the PFA is uncontrolled with unknown
validity limits makes this approach highly problematic.

Therefore, a technique is needed that facilitates Casimir
computations from field-theoretic first principles.  For this purpose,
{\em worldline numerics} has been developed \cite{Gies:2001zp},
combining the string-inspired approach to quantum field theory
\cite{Schubert:2001he} 
with Monte Carlo methods. As a main advantage, the worldline algorithm
can be formulated for arbitrary geometries, resulting in a numerical
estimate of the exact answer \cite{Gies:2003cv}.  For the 
sphere-plate and cylinder-plate configurations, also new
analytic methods are currently developed and latest results including
exact solutions are given in \cite{Bulgac:2005ku,Emig:2006}. In either
case, quantitatively accurate results for the experimentally relevant
parameter ranges are missing so far.

In this Letter, we use worldline numerics
\cite{Gies:2001zp,Gies:2003cv} to examine the Casimir effect in a
sphere-plate and cylinder-plate geometry for a fluctuating scalar
field, obeying Dirichlet boundary conditions (``Dirichlet
scalar''). We compute the Casimir interaction energies
that give rise to forces between the rigid surfaces. Thereby, we
quantitatively determine validity bounds for the PFA.  Apart from
numerical discretization, for which a careful error management on the
0.1\% level is performed, no quantum-field-theoretic approximation is
needed.

We emphasize that the Casimir energies for the Dirichlet scalar should
not be taken as an estimate for those for the electromagnetic (EM)
field, leaving especially the sphere-plate case as a pressing open
problem. Nevertheless, the validity constraints that we derive for the
PFA hold independently of that, since the PFA approach makes no
reference to the nature of the fluctuating field. If an experiment is
performed outside the PFA validity ranges determined below, any
comparison of the data with theory using the PFA has no firm basis.


\noindent
{\em Casimir curvature effects.} -- An intriguing property of the
Casimir effect has always been its geometry dependence. As long as the
typical curvature radii $R_i$ of the surfaces are large compared to
the surface separation $a$, the PFA is assumed to provide
for a good approximation. In this approach, the curved surfaces are
viewed as a superposition of infinitesimal parallel plates
\cite{pft1,Bordag:2001qi}. The Casimir interaction energy is obtained
by an integration of the parallel-plate energy applied to the
infinitesimal elements. Part of the curvature effect is introduced by
the choice of a suitable integration measure which is generally
ambiguous, as discussed, e.g., in \cite{Scardicchio:2004fy}.  For the
case of a sphere with radius $R$ at a (minimal) distance $a$ from a
plate, the PFA result at next-to-leading order reads
\begin{eqnarray}
\text{}\!\!\!E_{\text{PFA}}(a,R)\!
&=&\! E_{\text{PFA}}^{(0)}(a,R)\left(\! 1-
  \genfrac{\{}{\}}{0pt}{}{1}{3} \frac{a}{R}  
  + \mathcal O ((\tfrac{a}{R})^2)\!\! \right)\!,
  \label{eq:Enorm} \\
&&E_{\text{PFA}}^{(0)}(a,R)=-c_{\text{PP}}\frac{\pi^3}{1440}
\frac{R}{a^2}, \label{eq:Ezero}
\end{eqnarray}
where the upper (lower) coefficient in braces holds for the so-called
plate-based (sphere-based) PFA. They represent two limiting cases of
the PFA and have often been assumed to span the error bars for the
true result. Furthermore, $c_{\text{PP}}=2$ for an 
EM field or a complex scalar, and $c_{\text{PP}}=1$ for real scalar
field fluctuation.

Heuristically, the PFA is in contradiction with Heisenberg's
uncertainty principle, since the quantum fluctuations are assumed to
probe the surfaces only locally at each infinitesimal element. However,
fluctuations are not localizable, but at least probe the surface in a
whole neighborhood. In this manner, the curvature information
enters the fluctuation spectrum.  This quantum mechanism is
immediately visible in the worldline formulation of the Casimir
problem. Therein, the sum over fluctuations is
mapped onto a Feynman path integral. Each path (worldline) can be
viewed as a random spacetime trajectory of a quantum
fluctuation. Owing to a generic spatial extent of the worldlines, the
path integral directly samples the curvature properties of the
surfaces \cite{Gies:2003cv}.

\begin{figure}
\includegraphics[width=\columnwidth]{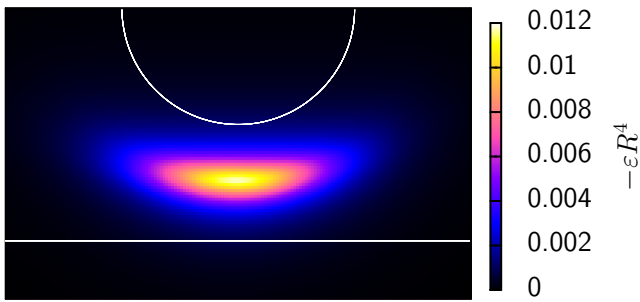}
\vspace{-.7cm}
\caption{Contour plot of the negative Casimir interaction energy
  density for a sphere of radius $R$ above an infinite plate; the
  sphere-plate separation $a$ has been chosen as $a=R$ here. The plot
  results from a pointwise evaluation of Eq.~\eqref{eq:ECasW} using
  worldlines with a common center of mass.}
\label{fig:contour}
\end{figure}

For the Dirichlet scalar, the worldline representation of the Casimir
interaction energy boils down to \cite{Gies:2003cv,Gies:2005ym}
\begin{equation}
E_{\text{Casimir}}=-\frac{1}{2} \frac{1}{(4\pi)^2} \int_{0}^\infty
\frac{d T}{T^3}\, e^{-m^2 T}\,\left\langle\Theta_\Sigma[x]
\right\rangle_x . \label{eq:ECasW} 
\end{equation}
The expectation value in \eqref{eq:ECasW} has to be taken with respect
to an ensemble of closed worldlines,
\begin{equation}
\langle \dots \rangle_x := \int_{x(T)=x(0)} \mathcal D x \, \dots
e^{-\frac{1}{4} \int_0^T d \tau {\dot x}^2},
\label{eq:VEV2}
\end{equation}
with implicit normalization $\langle 1 \rangle_x=1$. In
Eq.~\eqref{eq:ECasW}, $\Theta_\Sigma[x]=1$ if a worldline $x$
intersects both surfaces $\Sigma=\Sigma_1+\Sigma_2$, and
$\Theta_\Sigma[x]=0$ otherwise. A worldline with $\Theta_\Sigma[x]=1$
represents a boundary-condition violating fluctuation. Its removal
from the set of admissible fluctuations contributes to the negative
Casimir interaction energy.

We evaluate the worldline integral with Monte Carlo techniques,
generating an ensemble of $n_{\text{L}}$ worldlines with the v loop
algorithm \cite{Gies:2003cv}. Each worldline is characterized by $N$
points after discretizing its propertime. In this work, we
have used ensembles with up to $n_{\text{L}}=2.5\cdot10^5$ and
$N=4\cdot10^6$.
Details of the algorithmic improvements used for this work will be
given elsewhere \cite{HGKK}.

Further advanced field-theoretic methods have been developed for
Casimir calculations during the past years.  Significant improvements
compared to the PFA have been achieved by the semiclassical
approximation \cite{semicl}, a functional-integral approach using
boundary auxiliary fields \cite{Golestanian:1998bx}, and the optical
approximation \cite{Scardicchio:2004fy}. These methods are especially
useful for analyzing particular geometries by purely or partly
analytical means; in the general case, approximations are often
necessary but difficult to control. Hence, our results can also shed
light on the quality of such approximations.

\begin{figure}
\includegraphics[width=\columnwidth]{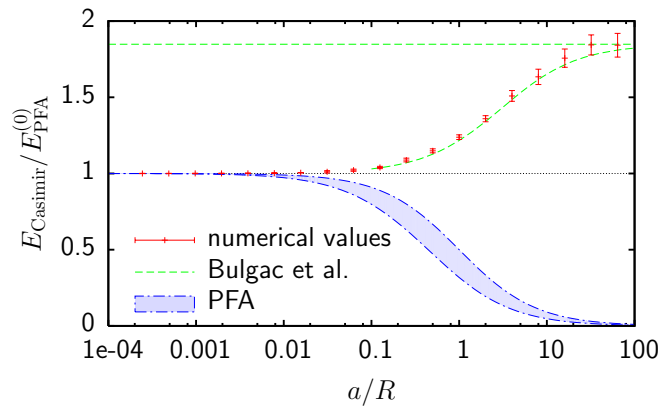}
\vspace{-.7cm}
\caption{Casimir interaction energy of a sphere with radius $R$ and an
infinite plate vs. the curvature parameter $a/R$. The energy is
normalized to the zeroth-order PFA formula (\ref{eq:Ezero}),
$E^{(0)}_{\text{PFA}}$. For larger curvature parameter, the PFA
estimate (dot-dashed line) differs qualitatively from the worldline
result (crosses with error bars). Here, we observe good agreement of
our result with the exact solution of \cite{Bulgac:2005ku} which is
available for $a/R\gtrsim 0.1$ (dashed line).}
\label{fig:sphere}
\end{figure}

\noindent
{\em Sphere above plate.} -- We consider a sphere with radius
$R$ above an infinite plate at a (minimal) distance $a$.  A contour
plot of the energy density along a radial plane obtained by a
pointwise evaluation of Eq.~\eqref{eq:ECasW} is shown in
Fig. \ref{fig:contour}. This density is related to the density of
worldlines with a given center-of-mass that intersect both surfaces.
Figure \ref{fig:sphere} presents a global view on the Casimir
interaction energy for a wide range of the curvature parameter $a/R$;
the energy is normalized to the zeroth order of the PFA formula
(\ref{eq:Ezero}), $E^{(0)}_{\text{PFA}}$.
For small $a/R$ (``large spheres''), our worldline result (crosses
with error bars) and the full sphere- and plate-based PFA estimates
(dashed-dotted lines) show reasonable agreement, settling at the
zeroth-order PFA $E^{(0)}_{\text{PFA}}$. The first field-theoretic
confirmation of this result has been obtained within the
semi-classical approximation in \cite{semicl}. The full PFA departs on
the percent level from $E^{(0)}_{\text{PFA}}$ for $a/R \gtrsim 0.01$,
exhibiting a relative energy decrease. By contrast, our worldline
result first stays close to $E^{(0)}_{\text{PFA}}$ and then increases
towards larger energy values relative to $E^{(0)}_{\text{PFA}}$. This
observation confirms earlier worldline studies \cite{Gies:2003cv} and
agrees with the optical approximation \cite{Scardicchio:2004fy} in
this curvature regime.

For larger curvature $a/R\gtrsim 0.1$ (``smaller spheres''),
we observe a strong increase relative to $E^{(0)}_{\text{PFA}}$
\cite{Gies:2005ym}. Here, our data satisfactorily agrees with the
exact solution found recently for this regime \cite{Bulgac:2005ku}
(dashed line). The latter work also provides for an exact asymptotic
limit for $a/R\to\infty$, resulting in $180/\pi^4$ for our
normalization.  Our worldline data confirms this limit in
Fig.  \ref{fig:sphere}.

Two important lessons can be learned from this plot: first, the PFA
already fails to predict the correct sign of the curvature effects
beyond zeroth order, see also \cite{Brevik:2004uw}. Second, the
relation between the Casimir effect for Dirichlet scalars and that for
the EM field is strongly geometry dependent. For the parallel-plate
case, Casimir forces only differ by the number of degrees of freedom,
cf. the coefficient $c_{\text{PP}}$ in Eq.~\eqref{eq:Enorm}. For large
curvature, the Casimir energy for the Dirichlet scalar scales with
$a^{-2}$, whereas that for the EM field obeys the Casimir-Polder law
$\sim a^{-4}$ \cite{Casmir:1947hx,DeKieviet}. We emphasize
that this difference does not affect our conclusions about the
validity limits of the PFA, because the PFA makes no reference to the
fluctuating field other than the coefficient $c_{\text{PP}}$.

\begin{figure}
\includegraphics[width=\columnwidth]{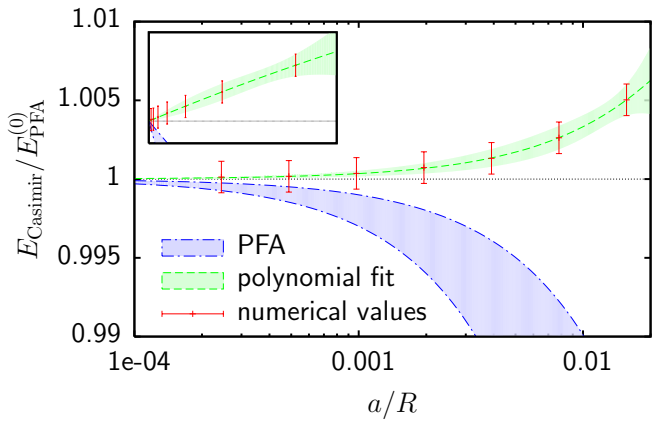}
\vspace{-.7cm}
\caption{Magnified view of Fig. \ref{fig:sphere} for small $a/R$. The
  0.1\% validity range of the PFA is characterized by curvature
  parameters, where the error band of our worldline results and the
  PFA band (blue-shaded/in between the dot-dashed lines) overlap, see
  Eq. \eqref{eq:acconepermille}. The dashed lines depict a constraint
  polynomial fit of the worldline result,
  $p(a/R)=1+0.35(a/R)-1.92(a/R)^2$, and its standard deviation. The
  inlay displays the same curves with a linear $a/R$ axis}
\label{fig:sphere_small_a}
\end{figure}
For a quantitative determination of the PFA validity limits,
Fig. \ref{fig:sphere_small_a} displays the zeroth-order normalized
energy for small curvature parameter $a/R$. Here, our result has an
accuracy of 0.1\% (jack-knife analysis). The error is dominated by the
Monte Carlo sampling and the ordinary-integration
accuracy; the error from the worldline discretization is found
negligible in this regime, implying a sufficient proximity to the
continuum limit.

In addition to our numerical error band, we consider the region
between the sphere- and the plate-based PFA as the PFA error band.  We
identify the 0.1\% accuracy limit of the PFA with the curvature
parameter $a/R|_{0.1\%}$ where the two bands do no longer
overlap. We obtain
\begin{equation}
\frac{a}{R}\Big|^{\text{PFA}}_{0.1\%}\leq 0.00073
\label{eq:acconepermille}
\end{equation}
as the corresponding validity range for the curvature parameter.  For
instance, for a typical sphere with $R=200\mu$m and an experimental
accuracy goal of $0.1\%$, the PFA should not be used for $a\gtrsim
150$nm. We conclude that the PFA should be dropped from the analysis
of future experiments.

For the 1\% accuracy limit of the PFA, we increase the band of our
worldline estimate by this size and again determine the curvature
parameter for which there is no intersection with the PFA
band anymore. We obtain
\begin{equation}
\frac{a}{R}\Big|^{\text{PFA}}_{1\%}\leq 0.00755.\label{eq:acconepercent} 
\end{equation}
For a sphere with $R=200\mu$m and an experimental accuracy goal of
$1\%$, the PFA holds for $a<1.5\mu$m. This result confirms the use of
the PFA for the data analysis of the corresponding experiments
performed so far.  

In order to study the asymptotic expansion of the normalized energy,
we fit our data to a second-order polynomial for $a/R<0.1$ and include
the exactly known result for $a/R\to 0$. We obtain $p(x)\simeq 1+ 0.35
x -1.92 x^2 \pm 0.19 x \sqrt{ 1 - 137.2 x + 5125 x^2}$, where
$x=a/R$. The fit result is plotted in Fig. \ref{fig:sphere_small_a}
(dashed lines), which illustrates that $E\simeq E^{(0)}_{\text{PFA}}\,
p(a/R)$ is a satisfactory approximation to the Casimir energy for
$a/R<0.1$, replacing the PFA \eqref{eq:Enorm}. The inlay in this
figure displays the same curves with a linear $a/R$ axis, illustrating
that the lowest-order curvature effect is linear in $a/R$. Given the
results of the PFA \eqref{eq:Enorm}, the semiclassical approximation
\cite{semicl}, $p_{sc}(x)\simeq1-0.17x$, cf. \cite{Bulgac:2005ku}, and
the optical approximation \cite{Scardicchio:2004fy},
$p_{opt}(x)\simeq1+0.05x$, the latter appears to estimate curvature
effects more appropriately.

\begin{figure}
\includegraphics[width=\columnwidth]{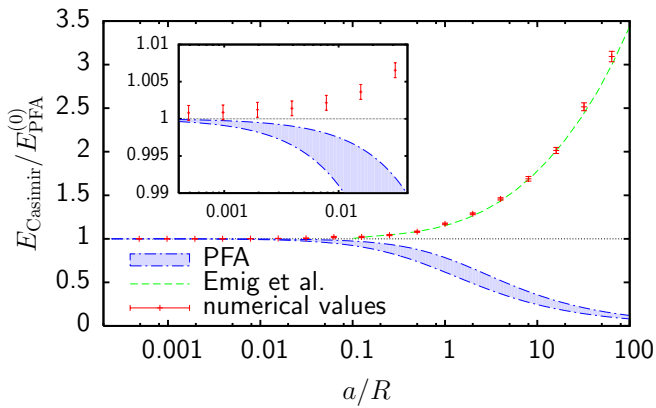}
\vspace{-.7cm}
\caption{Casimir interaction energy (normalized to
  $E_{\text{PFA}}^{(0)}$) of an infinitely long cylinder with radius
  $R$ at a distance $a$ above an infinite plate vs. the curvature
  parameter $a/R$. The inlay shows a magnified view for small values of
  $a/R$.}
\label{fig:cylinder}
\end{figure}

\noindent
{\em Cylinder above plate.} -- The cylinder-plate configuration is a
promising tool for high-precision experiments
\cite{Brown-Hayes:2005uf}, since the force signal increases linearly
with the cylinder length.  Figure \ref{fig:cylinder} shows the
corresponding Casimir interaction energy versus the curvature
parameter. The energy axis is again normalized to the zeroth-order PFA
result, $E_{\text{PFA}}^{(0)}(a,R)= -c_{\text{PP}}
\frac{3\pi}{4\sqrt{2}}\,\, \frac{R^{1/2}}{a^{5/2}}$.  The qualitative
conclusions for the validity of the PFA are similar to that for the
sphere above a plate: beyond leading order, the PFA even predicts the
wrong sign of the curvature effects. Quantitatively, the PFA validity
limits are a factor $\sim3$ larger than 
Eqs. \eqref{eq:acconepermille},\eqref{eq:acconepercent}, owing to the
absence of curvature along the cylinder axis.

The most important difference to the sphere-plate case arises for
large $a/R$. Here, the data is compatible with a log-like increase
relative to $E_{\text{PFA}}^{(0)}$, implying a surprisingly weak
decrease of the Casimir force for large curvature $a/R\to \infty$.
Our result agrees nicely with the very recent exact result
\cite{Emig:2006} which is available for $a/R\gtrsim0.1$. The data thus
confirms the observation of \cite{Emig:2006} that the resulting
Casimir force has the weakest possible decay, $F\sim 1/[a^3\ln(a/R)]$,
for asymptotically large curvature parameter $a/R\to \infty$.

In summary, we have computed Casimir interaction energies for the
sphere-plate and cylinder-plate configuration with Dirichlet boundary
conditions from first principles for a wide range of curvature
parameters $a/R$.  In general, we observe that curvature effects and
geometry dependencies are intriguingly rich, implying that naive
estimates can easily be misguiding. In particular, predictions based
on the PFA are only reliable in the asymptotic no-curvature limit. Its
quantitative validity bounds given above and thus genuine Casimir
curvature effects are in reach of currently planned experiments.

Beyond the Dirichlet scalar investigated here, it is well possible,
e.g., for the EM field, that some cancellation of curvature effects occurs
between modes obeying different boundary conditions. In fact, such a
partial cancellation between TE and TM modes of the separable
cylinder-plate geometry can be observed in the recent exact result for
the EM field for small curvature \cite{Emig:2006}. Casimir
calculations for the EM field in non-separable geometries, such as the
important sphere-plate case, therefore remain a prominent open
problem.

The authors are grateful to T.~Emig, R.L.~Jaffe, A.~Scardicchio, and
A.~Wirzba for useful discussions.  The authors acknowledge support by
the DFG under contract Gi 328/1-3 (Emmy-Noether program) and Gi
328/3-2.


\begin{thebibliography}{99}


\bibitem{Casimir:dh}
H.B.G.~Casimir,
Kon.\ Ned.\ Akad.\ Wetensch.\ Proc.\  {\bf 51}, 793 (1948).

\bibitem{Lamoreaux:1996wh}
S.~K.~Lamoreaux,
Phys.\ Rev.\ Lett.\  {\bf 78}, 5 (1997).

\bibitem{Mohideen:1998iz}
U.~Mohideen and A.~Roy,
Phys.\ Rev.\ Lett.\  {\bf 81}, 4549 (1998);

\bibitem{Roy:1999dx}
A.~Roy, C.~Y.~Lin and U.~Mohideen,
Phys.\ Rev.\ D {\bf 60}, 111101 (1999).

\bibitem{Ederth:2000}
T.~Ederth, Phys.\ Rev.\  A {\bf 62}, 062104 (2000)

\bibitem{Chan:2001}
H.B.~Chan, V.A.~Aksyuk, R.N.~Kleiman, D.J.~Bishop and F.~Capasso,
Science 291, 1941 (2001).

\bibitem{Bressi:2002fr}
G.~Bressi, G.~Carugno, R.~Onofrio and G.~Ruoso,
Phys.\ Rev.\ Lett.\  {\bf 88}, 041804 (2002).

\bibitem{Chen:2002}
F.~Chen, U.~Mohideen, G.L.~Klimchitskaya and V.M.~Mostepanenko,
Phys.\ Rev.\ Lett.\ {\bf 88}, 101801 (2002).

\bibitem{Bordag:1998nv}
M.~Bordag, B.~Geyer, G.~L.~Klimchitskaya and V.~M.~Mostepanenko,
Phys.\ Rev.\ D {\bf 58}, 075003 (1998), 
{\em ibid.} {\bf 60}, 055004 (1999), 
{\em ibid.} {\bf 62}, 011701 (2000).

\bibitem{Long:1998dk}
J.~C.~Long, H.~W.~Chan and J.~C.~Price,
Nucl.\ Phys.\ B {\bf 539}, 23 (1999).


\bibitem{Mostepanenko:2001fx}
V.~M.~Mostepanenko and M.~Novello,
Phys.\ Rev.\ D {\bf 63}, 115003 (2001).


\bibitem{Milton:2001np}
K.~A.~Milton, R.~Kantowski, C.~Kao and Y.~Wang,
Mod.\ Phys.\ Lett.\ A {\bf 16}, 2281 (2001).


\bibitem{Decca:2003td}
R.S.~Decca, E.~Fischbach, G.L.~Klimchitskaya, D.E.~Krause, D.L.~Lopez and V.M.~Mostepanenko,
Phys.\ Rev.\ D {\bf 68}, 116003 (2003), 
R.S.~Decca, D.~Lopez, H.B.~Chan, E.~Fischbach, D.E. Krause and C.R.~Jamell,
Phys.\ Rev.\ Lett.\  {\bf 94}, 240401 (2005).

\bibitem{Klimtchiskaya:1999}
G.L.~Klimchitskaya, A.~Roy, U.~Mohideen, and V.M.~Mostepanenko, Phys.\
Rev.\ A {\bf 60}, 3487 (1999).

\bibitem{Lambrecht:1999vd}
A.~Lambrecht and S.~Reynaud,
Eur.\ Phys.\ J.\ D {\bf 8}, 309 (2000).

\bibitem{Bezerra:2000} 
V.B.~Bezerra, G.L.~Klimchitskaya, and V.M.~Mostepanenko,
Phys. Rev. A 62, 014102 (2000).

\bibitem{Bordag:2001qi}
M.~Bordag, U.~Mohideen and V.~M.~Mostepanenko,
Phys.\ Rept.\  {\bf 353}, 1 (2001).

\bibitem{Graham:2002xq}
N.~Graham, R.~L.~Jaffe, V.~Khemani, M.~Quandt, M. Scandurra and H.~Weigel,
Nucl.\ Phys.\ B {\bf 645}, 49 (2002).

\bibitem{pft1}
B.V.~Derjaguin, I.I.~Abrikosova, E.M.~Lifshitz, Q.Rev. {\bf 10}, 295
(1956); 
J.~Blocki, J.~Randrup, W.J.~Swiatecki, C.F.~Tsang, Ann.~Phys.~(N.Y.)
{\bf 105}, 427 (1977).

\bibitem{Gies:2001zp}
H.~Gies and K.~Langfeld,
Nucl.\ Phys.\ B {\bf 613}, 353 (2001); 
Int.\ J.\ Mod.\ Phys.\ A {\bf 17}, 966 (2002).

\bibitem{Schubert:2001he}
see, e.g., C.~Schubert,
Phys.\ Rept.\  {\bf 355}, 73 (2001).

\bibitem{Gies:2003cv}
H.~Gies, K.~Langfeld and L.~Moyaerts,
JHEP {\bf 0306}, 018 (2003); 
arXiv:hep-th/0311168.

\bibitem{Bulgac:2005ku}
A.~Bulgac, P.~Magierski and A.~Wirzba,
arXiv:hep-th/0511056; 
A.~Wirzba, A.~Bulgac and P.~Magierski,
arXiv:quant-ph/0511057.

\bibitem{Emig:2006}
T.~Emig, R.~L. Jaffe, M.~Kardar, A.~Scardicchio, 
cond-mat/0601055.

\bibitem{Scardicchio:2004fy}
A.~Scardicchio and R.~L.~Jaffe,
Nucl.\ Phys.\ B {\bf 704}, 552 (2005);
Phys.\ Rev.\ Lett.\  {\bf 92}, 070402 (2004).


\bibitem{Gies:2005ym}
H.~Gies and K.~Klingm\"uller,
arXiv:hep-th/0511092.

\bibitem{HGKK} H.~Gies and K.~Klingm\"uller, in preparation.


\bibitem{semicl}
M. Schaden and L. Spruch, Phys.~Rev.~A {\bf 58}, 935 (1998);
Phys. Rev. Lett. {\bf 84} 459 (2000) 

\bibitem{Golestanian:1998bx}
R.~Golestanian and M.~Kardar, Phys.~Rev.~A {\bf 58}, 1713 (1998);
T.~Emig, A.~Hanke and M.~Kardar,
Phys.\ Rev.\ Lett.\  {\bf 87} (2001) 260402;
T.~Emig and R.~Buscher,
Nucl.\ Phys.\ B {\bf 696}, 468 (2004).

\bibitem{Brevik:2004uw}
I.~Brevik, E.K.~Dahl and G.O.~Myhr,
J.\ Phys.\ A {\bf 38}, L49 (2005).

\bibitem{Brown-Hayes:2005uf}
M.~Brown-Hayes, D.A.R.~Dalvit, F.D.~Mazzitelli, W.J. Kim and R.~Onofrio,
Phys.\ Rev.\ A {\bf 72}, 052102 (2005).

\bibitem{Casmir:1947hx}
H.B.G.~Casmir and D.~Polder,
Phys.\ Rev.\  {\bf 73}, 360 (1948).

\bibitem{DeKieviet}
V.~Druzhinina and M.~DeKieviet, Phys.~Rev.~Lett.~ {\bf  91}, 193202
(2003). 



\end{thebibliography}
\end{document}